\documentclass{eas}
\usepackage{graphicx}
\newcommand{\be}{\begin{equation}} \newcommand{\ee}{\end{equation}}
\newcommand{\bea}{\begin{eqnarray}} \newcommand{\eea}{\end{eqnarray}}

\newcommand{\re}[1]{(\ref{#1})}

\newcommand{\pat}{\partial}

\newcommand{\fig}[1]{figure \ref{#1}}
\newcommand{\brt}[1]{[#1]}

\newcommand{\LCDM}{$\Lambda$CDM\ }
\newcommand{\GN}{G_{\mathrm{N}}}
\newcommand{\ha}{\frac{1}{2}}

\newcommand{\keq}{k_{\mathrm{eq}}}

\newcommand{\teq}{t_{\mathrm{eq}}}

\newcommand{\rmd}{\mathrm{d}}

\newcommand{\adot}{\dot{a}}
\newcommand{\addot}{\ddot{a}}
\newcommand{\rhodot}{\dot{\rho}}

\newcommand{\HH}{\frac{\adot^2}{a^2}}

\newcommand{\av}[1]{\langle{#1}\rangle}
\newcommand{\sQ}{\mathcal{Q}}
\newcommand{\sR}{{^{(3)}R}}

\newcommand{\om}{\omega_{\mathrm{m}}}

\newcommand{\PRD}[1]{{\it Phys. Rev.} {\bf D#1}}

\newcommand{\MNRAS}[1]{{\it Mon. Not. Roy. Astron. Soc.} {\bf #1}}
\newcommand{\APJ}[1]{{\it Astrophys. J.} {\bf #1}}

\newcommand{\CQG}[1]{{\it Class. Quant. Grav.} {\bf #1}}
\newcommand{\GRG}[1]{{\it Gen. Rel. Grav.} {\bf #1}}

\newcommand{\PROG}[1]{{\it Prog. Theor. Phys.} {\bf #1}}

\newcommand{\IJMPD}[1]{{\it Int. J. Mod. Phys.} {\bf D#1}}

\begin{document}

\title{Structure formation as an alternative \\ to dark energy and modified gravity}

\runningtitle{Structure formation as an alternative to dark energy...}

\thanks{Based on a talk given at the CRAL-IPNL conference {\it Dark Energy and Dark Matter: Observations, Experiments and Theories}, on July 8, 2008.}
\author{Syksy R\"{a}s\"{a}nen}\address{Universit\'e de Gen\`eve, D\'epartement de Physique Th\'eorique, 24 quai Ernest-Ansermet, CH-1211 Gen\`eve 4, Switzerland; \email{syksy {\it dot} rasanen {\it at} iki {\it dot} fi}}
\begin{abstract}

Observations are inconsistent with a homogeneous and
iso-tropic universe with ordinary matter and gravity.
The universe is far from exact homogeneity and isotropy
at late times, and the effect of the non-linear
structures has to be quantified before concluding
that new physics is needed.
We explain how structure formation can lead to accelerated
expansion, and discuss a semi-realistic model where
the timescale of the change in the expansion rate emerges
from the physics of structure formation.

\end{abstract}

\maketitle

\section{Introduction}

\subsection{Three assumptions and a factor of 2}

The early universe, at least from one second onwards, is well
described by a model which is homogeneous and isotropic
(up to linear perturbations) and has only ordinary matter
and gravity. Here ordinary matter refers to matter with
non-negative pressure and ordinary gravity is based
on the four-dimensional Einstein-Hilbert action.
However, predictions of such models disagree with
observations of the late-time universe by a factor of two or so.
The angular diameter distance to the last scattering surface
at a redshift of 1100 measured from the cosmic microwave
background (CMB) is about 1.7 times smaller than
in the spatially flat homogeneous and isotropic model
dominated by pressureless matter.
(Assuming that the primordial perturbations
are adiabatic and that their spectrum is a power-law, and
normalising to the same value of the Hubble parameter today.)
Observations of type Ia supernovae and baryon
acoustic oscillations at redshifts of order one and below
are consistent with the CMB,
and indicate that the discrepancy arises at late times,
when the age of the universe is some billions of years.

In the context of the homogeneous and isotropic
Friedmann-Robertson-Walker (FRW) models, there
is a unique correspondence between the expansion rate and
the distance scale (given the spatial curvature), and
longer distances imply faster expansion. In terms of the
expansion rate, compared to expectations,
the Hubble parameter is a factor of 2 larger, given the
observed energy density of matter
($\Omega_\mathrm{m0}=8\pi\GN\rho_\mathrm{m0}/(3 H_0^2)\approx0.25$),
or a factor of 1.2--1.5 larger, given the age of the universe
($H_0 t_0\approx$ 0.8--1 instead of $H_0t_0=2/3$).

The factor 2 is observationally beyond reasonable doubt,
so at least one of the three assumptions of the model is wrong.
Either there is exotic matter with negative pressure
(``dark energy''), or
standard general relativity is not valid at large distances, or
the universe is not homogeneous and isotropic,
or some combination of the three.

There is no evidence for exotic matter or
modified gravity apart from observations of the
distance scale (and the related expansion rate).
For example, modifications of gravity have not been
observed in the solar system (apart from possibly
the Pioneer anomaly) (Reynaud \cite{Reynaud:2008}), and
dark energy has not been measured in the laboratory.
The situation is different from that of dark matter,
for which there are several independent lines of evidence,
such as the CMB peak structure, gravitational lensing,
the rotation curves of galaxies, the motions of clusters,
the early formation of structures and so on.
In contrast, the various measurements of the distance
scale (or the expansion rate) are probing the same
physical observable using different tracers.
An effect which increases the expansion rate (and
correspondingly the distance scale) at late times
can account for all of the observations.

Unlike the assumption of ordinary matter and gravity,
the assumption that the universe is homogeneous
and isotropy is known to be violated at late times,
when non-linear structure formation has started.
It is therefore necessary to quantify the effect of the
breakdown of homogeneity and isotropy on the expansion
rate before concluding that the observations call for
new physics.
It is suggestive that non-linear structures become significant at
late times, when large departures from the predictions
of the homogeneous and isotropic models with ordinary
matter and gravity are observed. In contrast,
models of exotic matter or modified gravity typically
have no natural scale corresponding to billions of years,
which is known as ``the coincidence problem''.

In section 2 we explain how inhomogeneity and/or
anisotropy can affect the expansion rate. In section
3 we discuss a toy model which helps to understand
the physics of acceleration due to structure formation.
In section 4 we discuss a semi-realistic model of
structures, and demonstrate how the correct timescale
of some billions of years emerges from the physics of structure
formation. In section 5 we conclude with a summary.
This is an overview of work reported in
(Rasanen \cite{Rasanen:2006a, Rasanen:2006b, Rasanen:2008a, Rasanen:2008b}).

\section{Backreaction}

\subsection{The fitting problem}

The effect of inhomogeneity and/or anisotropy on the
average expansion rate is known as backreaction.
To be specific, consider a manifold which one can split
into space and time. Take a fixed spatial domain
and consider average quantities such as the energy
density and the expansion rate in this domain.
If the domain is inhomogeneous and/or anisotropic,
the time evolution of these average quantities will 
be different than in a homogeneous and isotropic
space, even if the initial conditions are the same.

One might expect the effects of inhomogeneities
and/or anisotropies to be small
in the real universe, since the universe looks homogeneous
and isotropic when viewed on sufficiently large scales.
However, one has to distinguish between exact and
statistical homogeneity and isotropy.
The early universe is locally close to exact homogeneity
and isotropy, but after perturbations grow non-linear
and structure formation starts, the universe is only
statistically homogeneous and isotropic. In other words,
the local evolution is non-linear, but the distribution
of the non-linear regions is homogeneous and isotropic,
with a homogeneity scale which is today of the order 100 Mpc
(Hogg \cite{Hogg:2004}).
A space which is only statistically homogeneous and isotropic
expands differently than a locally homogeneous and isotropic
space, since averaging and time evolution do not commute.
This non-commutativity was first discussed in detail by George
Ellis (\cite{Ellis:1984}), and he termed the issue of finding
the model which best fits the average of the locally complex
behaviour ``the fitting problem''.

\subsection{The local equations}

It is possible to give some rather general results
on the average expansion rate.
Let us consider a general spacetime where the matter
can be treated as dust, that is, a pressureless ideal fluid.
After structure formation starts, this approximation does
not hold (Pueblas \& Scoccimarro \cite{Pueblas:2008}),
but the effects of velocity dispersion related to small scale
structures are not expected to be important for our discussion.
The geometry is determined by the Einstein equation
\bea \label{Einstein}
  G_{\alpha\beta} &=& 8 \pi \GN T_{\alpha\beta} = 8 \pi \GN \rho\, u_{\alpha} u_{\beta} \ ,
\eea

\noindent where $G_{\alpha\beta}$ is the Einstein tensor, $\GN$ is
Newton's constant, $T_{\alpha\beta}$ is the energy--momentum tensor,
$\rho$ is the energy density and $u^{\alpha}$
is the velocity of observers comoving with the dust.

We are interested in the average evolution of the geometry
determined by \re{Einstein}. In curved spacetime, only
scalars can be straightforwardly averaged, so we will
look at the scalar part of \re{Einstein}.
By taking the trace we obtain one scalar equation,
and projecting with $u^\alpha u^\beta$ and $u^\alpha \nabla^\beta$
gives two more (Buchert \cite{Buchert:1999}):
\bea
  \label{Rayloc} \dot{\theta} + \frac{1}{3} \theta^2 &=& - 4 \pi \GN \rho - 2 \sigma^2 + 2 \omega^2 \\
  \label{Hamloc} \frac{1}{3} \theta^2 &=& 8 \pi \GN \rho - \frac{1}{2} \sR + \sigma^2 - \omega^2 \\
  \label{consloc} \rhodot + \theta\rho &=& 0 \ ,
\eea

\noindent where a dot stands for derivative with respect to
proper time $t$ measured by observers comoving with the dust,
$\pat_t=u^\alpha\pat_\alpha$,
$\theta$ is the expansion rate of the local volume element,
$\sigma^2=\ha\sigma^{\alpha\beta}\sigma_{\alpha\beta}\geq0$
is the scalar built from the shear tensor $\sigma_{\alpha\beta}$,
$\omega^2=\ha\omega^{\alpha\beta}\omega_{\alpha\beta}\geq0$
is the scalar built from the vorticity tensor $\omega_{\alpha\beta}$,
and $\sR$ is the Ricci scalar on the tangent
space orthogonal to the fluid flow.
The equations are exact, and valid for arbitrary large variations
in density, expansion rate and other physical quantities.
The Hamiltonian constraint \re{Hamloc} is the analogue
of the Friedmann equation, and gives the local
expansion rate, and \re{consloc} expresses the
conservation of mass. The Raychaudhuri equation \re{Rayloc}
gives the local acceleration.
In the FRW case, we would have $\sigma^2=\omega^2=0$
and $\theta=3 H$, where $H$ is the Hubble parameter.

We take the vorticity to be zero. While rotation is
generated on small scales (Pueblas \& Scoccimarro \cite{Pueblas:2008}),
we expect that vorticity, like velocity dispersion, is not important
for the physics we discuss. Given $\omega^2=0$, we see from
\re{Rayloc} that the local acceleration is always negative
(or at most zero). One might be tempted to conclude that
since the local expansion decelerates everywhere at
all times, the average expansion must also decelerate.
However, this conclusion is false.

\subsection{The Buchert equations}

Let us see how the average expansion rate behaves
by averaging the exact equations \re{Rayloc}--\re{consloc}.
This was first done by Thomas Buchert (\cite{Buchert:1999}).
The average of a scalar quantity $f$ is simply its integral
over the spatial hypersurface of constant proper time,
divided by the volume,
\bea \label{av}
  \av{f}(t) \equiv \frac{ \int d^3 x \sqrt{^{(3)}g(t,\bar{x})} \, f(t,\bar{x}) }{ \int d^3 x \sqrt{^{(3)}g(t,\bar{x})} } \ ,
\eea

\noindent where $^{(3)}g$ is the determinant of the metric on the
hypersurface of constant proper time.

Averaging \re{Rayloc}--\re{consloc}, we obtain the Buchert equations
\bea
  \label{Ray} 3 \frac{\addot}{a} &=& - 4 \pi \GN \av{\rho} + \sQ \\
  \label{Ham} 3 \HH &=& 8 \pi \GN \av{\rho} - \frac{1}{2}\av{\sR} - \frac{1}{2}\sQ \\
  \label{cons} && \pat_t \av{\rho} + 3 \frac{\adot}{a} \av{\rho} = 0 \ ,
\eea

\noindent where the backreaction variable $\sQ$ contains the effect
of inhomogeneity and anisotropy:
\bea \label{Q}
  \sQ \equiv \frac{2}{3}\left( \av{\theta^2} - \av{\theta}^2 \right) - 2 \av{\sigma^2} \ ,
\eea

\noindent and we have defined a scale factor $a(t)$ such that the
volume of the spatial hypersurface is proportional to $a(t)^3$,
\bea  \label{a}
  a(t) \equiv \left( \frac{ \int d^3 x \sqrt{^{(3)}g(t,\bar{x})} }{ \int d^3 x \sqrt{^{(3)}g(t_0,\bar{x})} } \right)^{\frac{1}{3}}  \ ,
\eea

\noindent where $a$ has been normalised to unity at time $t_0$,
which we take to be today. As $\theta$ gives the expansion rate
of the volume, this definition of $a$ is equivalent to
$3\adot/a\equiv\av{\theta}$.
We will also use the notation $H\equiv\adot/a$.

The Buchert equations \re{Ray}--\re{cons} differ from the
FRW equations by the presence of the backreaction variable
$\sQ$ and the related fact that the spatial curvature is not
necessarily proportional to $a^{-2}$. Because of this
extra degree of freedom, the equations are not closed,
but they provide a constraint on the expansion history.
The backreaction variable $\sQ$ in \re{Q} has two parts. The shear
is also present in the local equations \re{Rayloc}--\re{Hamloc},
and acts to decelerate the expansion. The other term is
the variance of the expansion rate, which has no counterpart in the
local equations, and may in this sense be called emergent.
This term expresses the non-commutativity of time evolution
and averaging, and makes it possible for
the average equations to have qualitatively different
behaviour than the local equations. The contribution of the
variance to the acceleration is always positive (except in
the case of homogeneous expansion, when it is zero).
If the variance is large enough compared to the contribution
of the shear and the energy density, the average expansion rate
accelerates, as \re{Ray} indicates, even though locally the
expansion decelerates everywhere.

\section{Demonstrating acceleration}

\subsection{A two-region toy model}

The reason why the average expansion rate can increase
even though the local expansion decelerates is simple.
In an inhomogeneous space, different regions expand at different
rates, and the volume of faster expanding regions increases
more rapidly, by definition. Therefore the fraction of the
volume which is expanding faster increases, and the average
expansion rate can rise.

Structure formation in the universe starts from a density
distribution which is very smooth, with only small local
variations. In a simplified picture, overdense regions
slow down as their density contrast grows, and eventually
turn around and collapse to form stable structures.
Underdense regions in turn become ever emptier, and their
expansion rate increases (relative to the mean).

In order to clarify the physics of the acceleration due
to inhomogeneity and/or anisotropy, we will consider a
simple toy model of structure formation presented in 
(Rasanen \cite{Rasanen:2006a, Rasanen:2006b}), before
discussing a semi-realistic model in the next section.
We consider two spherical regions, one overdense and one underdense,
with scale factors denoted by $a_1$ and $a_2$, respectively.
In the Newtonian regime, a spherically symmetric region
expands like a FRW universe with the same mean density,
regardless of the density profile. This is known as the
spherical collapse model in the overdense case.

We take the underdense region, which models a cosmological void,
to be completely empty, so it expands like $a_1\propto t$.
The evolution of the overdense region, which models
the formation of a structure such as a cluster, is given by
$a_2\propto 1-\cos\phi$, $t\propto \phi-\sin\phi$, where
the parameter $\phi$ is called the development angle.
The value $\phi=0$ corresponds to the big bang singularity,
from which the overdense region expands until $\phi=\pi$,
when it turns around and starts collapsing. The region
shrinks to zero size at $\phi=2\pi$.
In studies of structure formation, the collapse
is usually taken to stabilise at $\phi=3\pi/2$
due to vorticity and velocity dispersion, and we will also
follow the evolution only up to that point.

The total volume in the toy model is $a^3=a_1^3+a_2^3$.
The average expansion rate and acceleration are
\bea
  \label{Hex} \!\!\!\!\!\!\!\!\!\!\!\!\!\!\!\!\!\!\!
H &=& \frac{ a_1^3 }{ a_1^3 + a_2^3 } H_1 + \frac{ a_2^3 }{ a_1^3 + a_2^3 } H_2 \equiv v_1 H_1 + v_2 H_2 \\
  \label{qex} \!\!\!\!\!\!\!\!\!\!\!\!\!\!\!\!\!\!\!
q &\equiv& - \frac{1}{H^2} \frac{\addot}{a} = \frac{H_1^2}{H^2} v_1 q_1 + \frac{H_2^2}{H^2} v_2 q_2 - 2 v_1 v_2 \frac{(H_1-H_2)^2}{H^2} \ ,
\eea

\noindent where we have expressed the acceleration in terms
of the deceleration parameter $q$ (note that negative $q$
corresponds to positive acceleration). The average expansion rate
is the volume-weighted average of the expansion rates $H_1$
and $H_2$, and it is therefore bounded from above by the fastest
local expansion rate.
In contrast, the acceleration is not simply the average of
the local accelerations $q_1$ and $q_2$.
In addition to the first two terms in \re{qex}, which 
are the local values, there is a third term, which is
non-positive, and contributes towards acceleration.
This term depends on the difference of the expansion rates of the
two regions, and corresponds to the variance part of the backreaction
variable $\sQ$ in \re{Ray}.

\begin{figure}
\hfill
\begin{minipage}[t!]{5.1cm} 
\scalebox{1.2}{\includegraphics[angle=0, clip=true, trim=0cm 2cm 0cm 0cm, width=\textwidth]{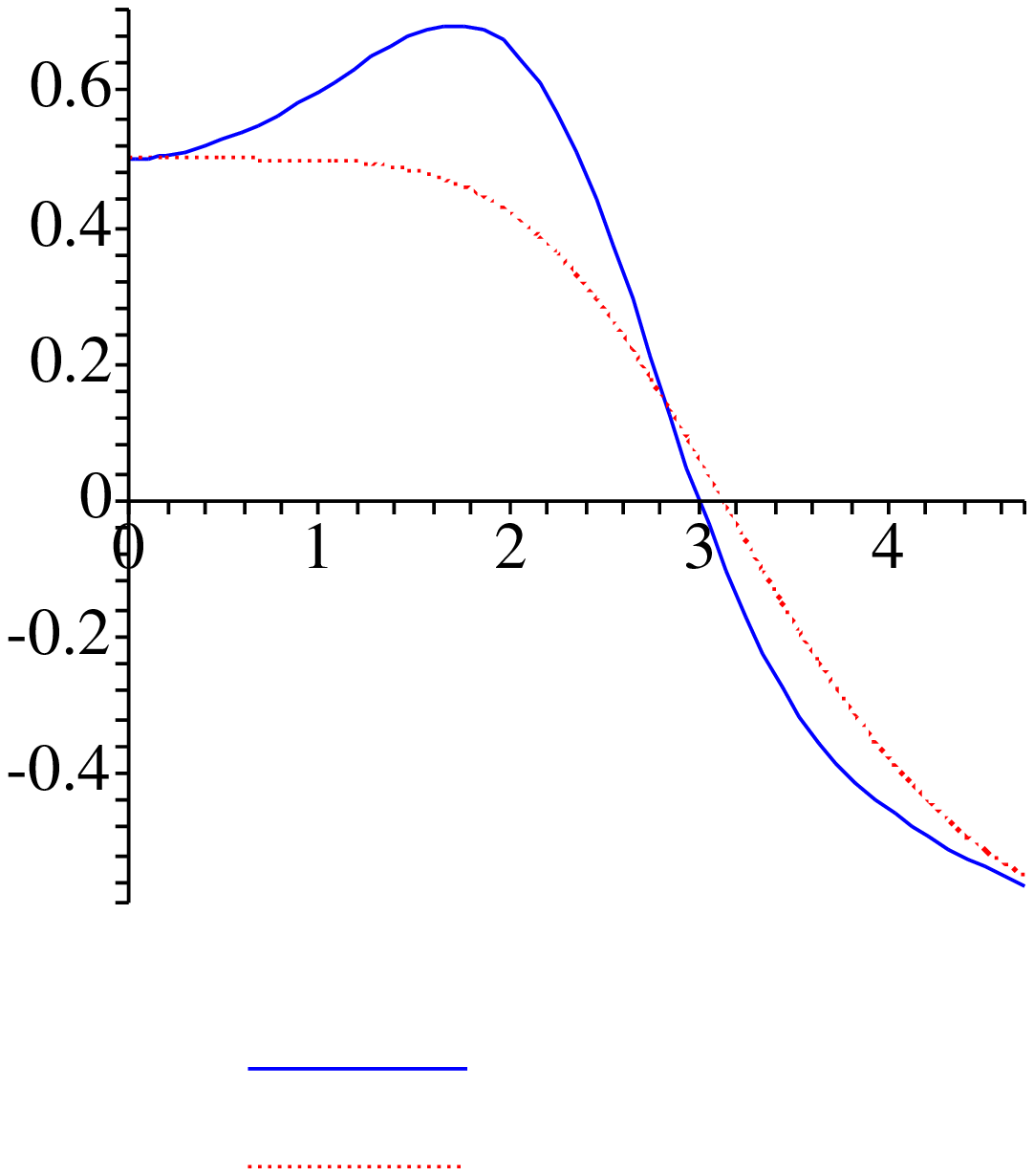}}
\begin{center} {\bf (a)} \end{center}
\end{minipage}
\hfill
\begin{minipage}[t!]{5.1cm}
\scalebox{1.2}{\includegraphics[angle=0, clip=true, trim=0cm 2cm 0cm 0cm, width=\textwidth]{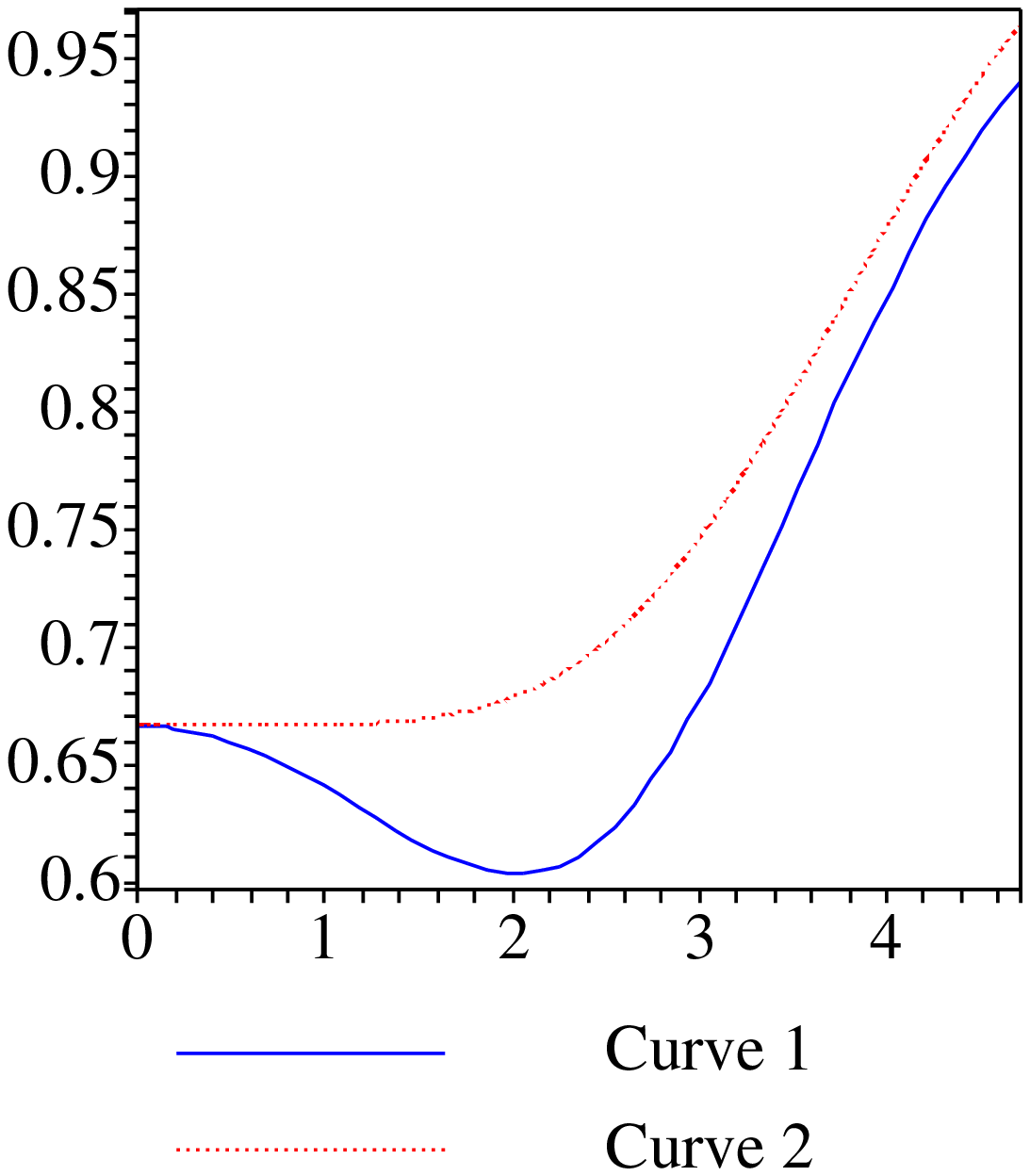}}
\begin{center} {\bf (b)} \end{center}
\end{minipage}
\hfill
\caption{The evolution of the toy model as a function of the
development angle $\phi$.
(a): The deceleration parameter $q$ in the toy model (blue, solid)
and in the \LCDM model (red, dash-dot).
(b): The Hubble parameter multiplied by time, $H t$, in the toy model
(blue, solid) and in the \LCDM model (red, dash-dot).}
\label{fig:toy}
\end{figure}

The toy model has one free parameter, the relative size of the
two regions at some time. We fix this by setting the
deceleration parameter $q$ to be equal to that of the
spatially flat \LCDM FRW model when $\Omega_\Lambda=0.7$.
In \fig{fig:toy} (a) we plot $q$ as a function
of the development angle $\phi$. In addition to the
toy model, we also show the \LCDM model
for comparison. This is done to help understand
the toy model, and does not imply that the toy model should
be taken seriously in a quantitative sense.

The \LCDM model starts matter-dominated, with $q=1/2$.
As vacuum energy becomes important, the model
decelerates less and then crosses over to acceleration.
Asymptotically, $q$ approaches $-1$ from above as the
Hubble parameter approaches a constant.
The toy model also starts with the FRW matter-dominated
behaviour. As the overdense region slows down, the
expansion decelerates more. Because the expansion of
the overdense region is slower, the underdense region
eventually takes over and comes to dominate the
expansion rate. The expansion then decelerates less,
and eventually accelerates; in fact the acceleration is
stronger than in the \LCDM model.

The acceleration is not due to regions speeding up
locally, but due to the slower region being less
represented in the average. This is particularly
easy to understand after the overdense region has started
collapsing at $\phi=\pi$. Then the contribution of the
overdense region $v_2 H_2$ to \re{Hex} is negative, but
the magnitude shrinks rapidly as $v_2$ decreases,
so it is transparent that the expansion rate increases.
Note that while there is an upper bound on the expansion
rate, there is no lower bound on the collapse rate.
Therefore, the acceleration can be arbitrarily rapid,
and $q$ can even reach minus infinity in a finite time.
(This simply means that the collapsing region
is so dominant that $H^2$ vanishes in the denominator of $q$.)
This is in contrast to FRW models, where $q\geq-1$
unless the null energy condition (or the modified gravity
equivalent) is violated.

Figure \ref{fig:toy} (b) shows the Hubble parameter
multiplied by time as a function of the development angle
$\phi$. This is the same information as in \fig{fig:toy} (a),
but plotted in terms of the first derivative of the scale
factor instead of the second derivative.
In the \LCDM model, $H t$ starts from $2/3$ in the
matter-dominated era and increases monotonically without
bound as $H$ approaches a constant. In the toy model,
$H t$ falls as the overdense region slows down, then rises
as the underdense region takes over, approaching $H t=1$
from below. The Hubble parameter in the toy model is at all times
smaller than in the \LCDM model, and because $H$ is bounded
from above by the fastest local expansion rate,
$H t$ cannot exceed unity.
This bound is valid also in realistic models:
as long as the matter can be treated as dust and vorticity
can be neglected, inhomogeneous and/or anisotropic models
predict $H t\leq1$, in contrast to FRW models with exotic
matter or modified gravity.
Observationally, $H t\approx$ 0.8--1.0 today
(see section 5.1 of (Rasanen \cite{Rasanen:2008a})).

Note that in an inhomogeneous and/or anisotropic dust model,
whether the expansion accelerates simply depends on how rapidly
faster expanding regions catch up with slower ones:
roughly speaking, how steeply the $H t$ curve rises.
This is why the variance in $\sQ$ \re{Q} contributes to
acceleration in the average Raychaudhuri equation \re{Ray}.
With a larger variance, the difference between slow
and fast regions is bigger, and faster
regions take over more rapidly.

\section{Towards reality}

\subsection{A semi-realistic model of backreaction}

The toy model shows that acceleration due to structure
formation is possible, and helps to understand what this
means physically. This possibility has also been
demonstrated with the exact spherically symmetric dust
solution, the Lema\^{\i}tre-Tolman-Bondi model
(Chuang \etal \ \cite{Chuang:2005}; Kai \etal \ \cite{Kai:2006};
Paranjape \& Singh \cite{Paranjape:2006}).

The issue is then whether this possibility is realised in
the universe. The non-linear evolution of structures
is too complex to follow exactly. However, because the
universe is statistically homogeneous and isotropic, knowing
some statistical properties is enough for evaluating
the average expansion rate.
In terms of the Buchert equations \re{Ray}--\re{cons},
the average expansion rate is completely determined
once the variance and the shear are known.
The average expansion rate is also determined if we
know which fraction of the universe is in which
state of expansion or collapse at each time.
We will now discuss a semi-realistic model which
does this by extending the two fixed regions of the
toy model to a continuous distribution of regions which evolves
in time (Rasanen \cite{Rasanen:2008a, Rasanen:2008b}).

The starting point is the spatially flat matter-dominated FRW
model, with a linear Gaussian field of density fluctuations.
Structures are identified with spherical peaks (and troughs)
of the density field smoothed on some scale, following
Bardeen (\cite{Bardeen:1986}). The peak number density as a function
of the smoothing scale and peak height can be determined
analytically. We keep the smoothing threshold fixed such
that $\sigma(t,R)=1$, where $\sigma$ is the root mean square
density contrast, $t$ is time and $R$ is the smoothing scale.
Non-linear structures typically form at $\sigma\approx1$, so
$R$ corresponds to the size of the typical largest structures,
and grows in time. The smoothing is just a simplified
treatment of the complex stabilisation and evolution of
structures in the process of hierarchical structure formation.
Since the peaks are spherical and isolated, and they are individually
assumed to be in the Newtonian regime, their expansion rate
is the same as that of a FRW universe with the same density,
as in the toy model. The fraction of volume which is not in
peaks is taken to expand like the spatially
flat matter-dominated FRW model.

The average expansion rate can be written as
\bea \label{H}
  H(t) = \int_{-\infty}^{\infty} \rmd\delta\, v_\delta(t) H_\delta(t) \ , 
\eea

\noindent where $v_\delta$ (or, more accurately, $\rmd v_\delta/\rmd\delta$)
is the fraction of the volume in regions with linear density
contrast $\delta$ and expansion rate $H_\delta(t)$.
The correspondence between $\delta$ and $H_\delta$ is given
by the spherical evolution model (i.e. the FRW evolution),
and the distribution of regions $v_\delta(t)$ is given by the
peak statistics, which is determined by the power spectrum
of the Gaussian density field. 
The power spectrum consists of the primordial spectrum,
determined in the early universe by some process such
as inflation, and the transfer function, which describes
the evolution of the modes at later times.

\begin{figure}[t]
\centering
\scalebox{0.5}
{\includegraphics[angle=0, clip=true, trim=0cm 0cm 0cm 0cm, width=\textwidth]{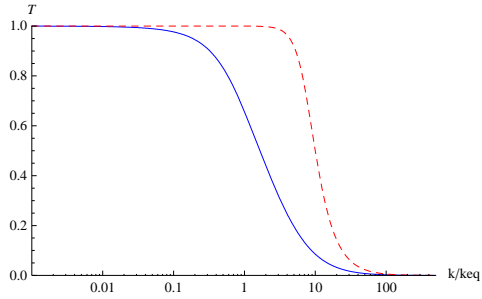}}
\caption{The BBKS (blue, solid) and BDG (red, dashed) transfer functions as a function of $k/\keq$.}
\label{fig:transfer}
\end{figure}

We take a scale-invariant primordial spectrum
and the cold dark matter transfer function.
For the latter, we consider two different
approximations in order to show the uncertainty
in the calculation. The BBKS transfer function
(Bardeen \cite{Bardeen:1986}) is a fit
to numerical calculations (we take a baryon fraction
of $0.2$), and the BDG form introduced by Bonvin \etal\ (\cite{Bonvin:2006})
is a simple analytically tractable function
with the correct qualitative features.
In \fig{fig:transfer} we show the suppression of the
amplitude of modes as a function of $k/\keq$,
the wavenumber divided by the wavenumber of the modes
which enter the horizon at matter-radiation equality.
The amplitude of modes which enter the horizon during the
radiation-dominated phase is suppressed, while modes entering
during matter domination retain their original amplitude.
The only scale present in the model is the one associated
with the turnover near matter-radiation equality.
With the transfer function fixed, there are no free
parameters: the expansion history $H(t)$ given by \re{H}
is completely fixed.

\begin{figure}
\hfill
\begin{minipage}[h]{6cm} 
\scalebox{1.0}{\includegraphics[angle=0, clip=true, trim=0cm 0cm 0cm 0cm, width=\textwidth]{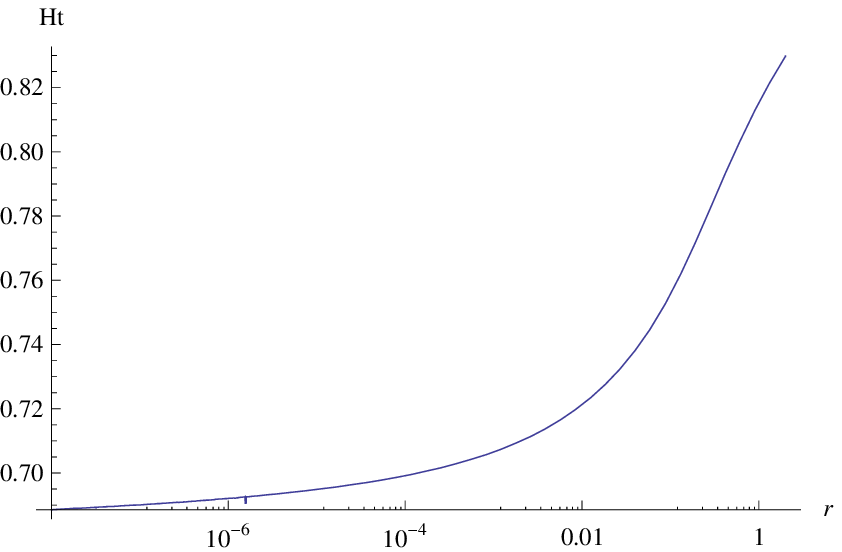}}
\begin{center} {\bf (a)} \end{center}
\end{minipage}
\hfill
\begin{minipage}[h]{6cm}
\scalebox{1.0}{\includegraphics[angle=0, clip=true, trim=0cm 0cm 0cm 0cm, width=\textwidth]{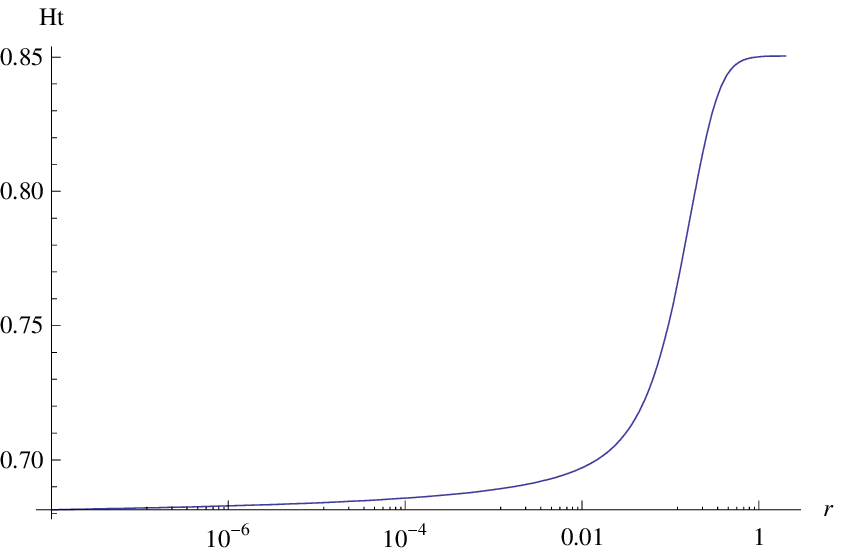}}
\begin{center} {\bf (b)} \end{center}
\end{minipage}
\hfill
\caption{The expansion rate $Ht$ as a function of $r=\keq R$ for
(a) the BBKS transfer function and (b) the BDG transfer function.}
\label{fig:Htr}
\end{figure}

In \fig{fig:Htr} we show $H t$ as a function of $r\equiv\keq R$,
the smoothing scale divided by the matter-radiation equality scale.
As in the toy model, $H t\approx 2/3$ at early times. As time goes
on, ever larger non-linear structures form, and they
take up a larger fraction of the volume.
The expansion rate grows (relative to the FRW value) slowly,
until there is a rapid rise and saturation, roughly at the scale
of matter-radiation equality.
It is clear that after $r=1$, when the perturbations which
correspond to the matter-radiation equality scale collapse,
$H t$ must settle to a constant, since the transfer function
is essentially unity, and there is no scale in the system anymore.

\begin{figure}
\hfill
\begin{minipage}[t]{6cm} 
\scalebox{1.0}{\includegraphics[angle=0, clip=true, trim=0cm 0cm 0cm 0cm, width=\textwidth]{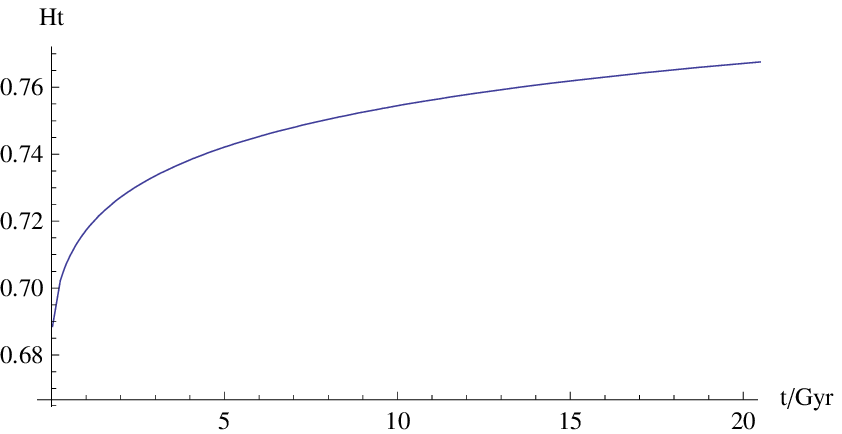}}
\begin{center} {\bf (a)} \end{center}
\end{minipage}
\hfill
\begin{minipage}[t]{6cm}
\scalebox{1.0}{\includegraphics[angle=0, clip=true, trim=0cm 0cm 0cm 0cm, width=\textwidth]{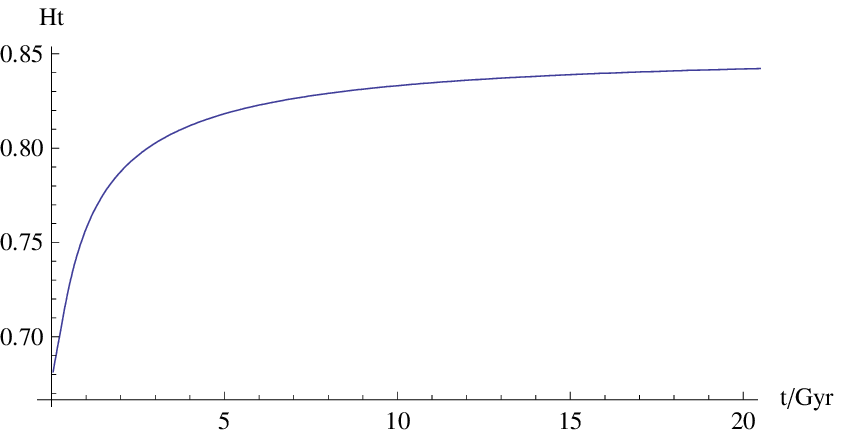}}
\begin{center} {\bf (b)} \end{center}
\end{minipage}
\hfill
\caption{The expansion rate $Ht$ as a function of time for
(a) the BBKS transfer function and (b) the BDG transfer function.}
\label{fig:Ht}
\end{figure}

The matter-radiation equality scale is
$\keq^{-1}\approx 13.7 \om^{-1}$ Mpc $\approx$ 100 Mpc,
using the (somewhat model-dependent) value $\om=0.14$
(Komatsu \cite{Komatsu:2008}). Observationally, $\sigma(t,R)\approx1$
today on scales somewhat smaller than 8 $h^{-1}$ Mpc, so
$R_0\approx$ 10 Mpc. Therefore the present day happens to be
located around $r=0.1$ in the plots - exactly in the transition region.

In \fig{fig:Ht} we show $H t$ as a function of time (in billions
of years) instead of the smoothing scale. The timescale is set
by the matter-radiation equality time $\teq$ and the amplitude
$A$ of the primordial density perturbations, $t\propto A^{-3/2}\teq$.
The matter-radiation equality time is $\teq\approx 1000 \om^{-2}$ years
$\approx$ 50 000 years for $\om=0.14$.
The primordial amplitude of the density perturbations inferred
from the observations depends on
the damping of the perturbations after deviations from the
matter-dominated FRW evolution become important. We take
the same amplitude as in the \LCDM model with $\Omega_\Lambda=0.7$
today, $A=4.6\times 10^{-5}$ (Dodelson \cite{Dodelson:2003}).
Ignoring damping by taking the matter-dominated FRW value
$A=1.9\times 10^{-5}$ would make the timescale longer by
a factor of $3.8$.
The slope of the $H t$ curve is less steep as a function
of time than as a function of $r$, because the size of
structures grows more slowly at late times.

The timing of the rise of $Ht$ is remarkably close to
the observed acceleration era, and the
change is of the right order of magnitude, 15--25\%.
Nothing related to the present time has been
used as an input in the model, the timing comes from
the slope of the transfer function and the primordial amplitude.

As noted when discussing the toy model, acceleration is a
quantitative question related to the slope of the $H t$ curve.
In the present case, while the expansion rate increases relative
to the FRW case, the rise is not sufficiently rapid to
correspond to acceleration.
This is related to the fact that, unlike in the toy model,
the overdense regions play almost no role, and
the evolution of $H t$ can be understood entirely in terms
of the underdense voids.
At early times, voids take up only a small fraction of the
volume, and $H t$ rises smoothly as they become prominent.
In order for the change to be more drastic, the expansion rate
should slow down due to overdense regions before the voids
take over, as in the toy model.

The model involves a number of approximations,
and it should not be trusted to more than an order of magnitude.
For example, we have not taken into account in the statistics
that peaks located inside larger peaks should not be double
counted, and we should also exclude troughs which are inside
peaks, while the reverse is not true (overdense regions can be
located inside voids), which breaks the symmetry of the
initial Gaussian density field (Sheth \& van de Weygaert \cite{Sheth:2003}).
Also, treating the peaks as isolated may not a
good approximation once the peak density is high,
collapsing structures are typically highly non-spherical,
even the more accurate BBKS transfer function has an error
of about 30\% and so on.
A more realistic calculation with quantified errors
is necessary to estimate the effect of structures
on the expansion rate in detail.

\section{Conclusions}

\subsection{Summary}

Observations are inconsistent with a
homogeneous and isotropic universe with ordinary matter
and gravity. FRW models do not include the effect of
non-linear structures on the expansion rate.
The Buchert equations, where the effect of structures
is taken into account, show that the average expansion of
a space can accelerate due to inhomogeneity and/or
anisotropy.

With a simple toy model involving one overdense and one
underdense region, we have demonstrated how acceleration
is possible. First the expansion rate slows down due to
the overdense region, and when the underdense region takes
over, the change is sufficiently rapid that the expansion
accelerates. The physical interpretation is simple: the
fraction of volume in the faster expanding region
grows, so the average expansion rate rises.

The important question is whether this happens in the
real universe. We have discussed a semi-realistic model
which involves an evolving distribution of regions, with
statistics given by the peak number density of the initial
Gaussian density field. In the model, the expansion rate
rises at some billions of years by about 15--25\%,
the right order of magnitude. The physical reason is that
the volume of the universe becomes dominated by voids.
The model has no acceleration, because overdense
regions are never prominent.
This is not crucial, since the model is probably not
correct beyond an order of magnitude, due to
various approximations, such as the spherical
symmetry of the structures and having equal amounts
of mass in the overdense and underdense regions.
It is remarkable that the correct timescale
emerges from the physics of structure formation, essentially
from the primordial amplitude of fluctuations
and the time of matter-radiation equality imprinted
on the cold dark matter transfer function. However, a
more realistic calculation is needed to determine
the details of the effect of structure formation on
the expansion rate.

One important aspect which we have not discussed here
is the relationship between the average expansion rate
and observations of distance and redshift. Physically,
it is clear that as the volume expands more rapidly,
distances will grow longer. However, the quantitative
relationship between the expansion rate and the distance
scale remains to be determined (this is work in progress).

That structure formation could account for the observations
of the late-time distance scale and expansion rate
is a plausible possibility.
Until the effect is quantified in detail, we do not know
whether new physics is needed or the observations
can be understood in terms of a complex realisation of
general relativity.

\end{document}